# Direct Nucleation of Hierarchical Nanostructures on Plasmonic Fiber Optics Enables Enhanced SERS Performance

*Di Zheng[1,2,\*,†], Riccardo Scarfiello[1,3,†], Muhammad Fayyaz Kashif[1,4], Liam Collard[1,5], Linda Piscopo[1,6], Maria Samuela Andriani[1,6], Elisabetta Perrone[3], Concetta Nobile[3], Massimo De Vittorio[1,6,7,\*], Ferruccio Pisanello[1,\*,†], Luigi Carbone[1,3,\*,†]*

[1] *Istituto Italiano di Tecnologia, Center for Biomolecular Nanotechnologies, Arnesano, Lecce, 73010, Italy*

[2] *State Key Laboratory of Radio Frequency Heterogeneous Integration, Shenzhen University, Shenzhen 518060, China*

[3] *CNR NANOTEC, Institute of Nanotechnology, Lecce, 73100, Italy*

[4] *Dipartimento di Ingegneria Elettrica e delle Tecnologie dell'Informazione, Università Degli Studi di Napoli Federico II, 80125 Napoli, Italy*

[5] *Comprehensive Cancer Centre, School of Cancer and Pharmaceutical Sciences, King's College London, London, UK*

[6] *Dipartimento di Ingegneria Dell'Innovazione, Università del Salento, Lecce 73100, Italy*

[7] *IDUN section, Department of Health Technology, Technical University of Denmark, DK-2800 Kgs. Lyngby, Denmark*

[\*] *Authors to whom correspondence should be addressed: dr.zhengdi@outlook.com, massimo.devittorio@iit.it, ferruccio.pisanello@iit.it, luigi.carbone@nanotec.cnr.it*

[†] *These authors contributed equally to this work*

## Abstract

We present an innovative fabrication method to achieve bottom-up in situ surface-overstructured Au nanoislands (NIs) with tunable grades of surface coverage, elongation, and branching, directly on micro-optical fibers for sensing applications. These all-in-gold hierarchical nanostructures consist of NIs coated with surface protrusions of various morphologies. They are created in solution using a selective seeded growth approach, whereby additional gold growth is achieved over Au NIs formerly developed on the fiber facet by a solid-state dewetting approach. The morphology of nanosized surface-NI overstructuring can be adjusted from multi-dot-decorated Au NIs to multi-arm-decorated Au NIs. This engineering of optical fibers allows for improved remote surface-enhanced Raman spectroscopy (SERS) molecular detection. By combining solid-state dewetting and wet-chemical approaches, we achieve stable in-contact deposition of surface-overstructured NIs with the optical fiber solid substrate, alongside precise control over branching morphology and anisotropy extent. The fiber optic probes engineered by surface-overstructured NIs exhibit outstanding sensing performance in an instant and through-fiber detection scheme, achieving a remarkable detection limit at $10^{-7}$ M for the R6G aqueous solution. These engineered probes demonstrate an improved detection limit by one order of magnitude and enhanced peak prominence compared to devices solely decorated with pristine NIs.

## Introduction

Enabling fiber optics probes with surface-enhanced Raman scattering (SERS) sensing ability can extend SERS applications to scenarios where sampling is challenging and minimal invasiveness to the system is required, such as probing in-vivo biological tissue[1,2], site-specific study for living cells[3,4], and on-site environmental monitoring[5–7]. Plasmonic nanoparticles are the most popular candidate to be integrated into the fiber tip since: (i) they support local plasmon resonances (LPRs), generating strong near-field enhancement[8–10], (ii) they can actively interact with guided modes through near-field coupling[11,12], (iii) bottom-up fabrication approaches compatible with both flat and curved fiber facets can enable wide-surface and high throughput fabrication [13–17]. A particularly interesting type of plasmonic nanoparticle in this context is the branched nanoparticle, an anisotropic nanostructure that exhibits multiple localized plasmon resonances (LPRs) with distinct resonant wavelengths, optimal performance at longer wavelengths and strong electromagnetic fields near the branches[18–20]. The increased surface area provides extended sites to interact with analytes, contributing to enhanced SERS performance compared with simple particle-shaped structures.

Efforts have been made to obtain multi-branched nanostructures (MBNSs) on the fiber facet to fabricate SERS-active fiber probes, and the currently employed methods are mainly based on electrostatic self-assembly (ESA). ESA requires first synthesis of the target batch of MBNSs with high purity, then using molecular linker to functionalize the silica or glass surface to obtain heterogeneous charges between the fiber and the MBNS' surfaces to realize nanoparticles immobilized on the fiber surface[21–23]. To circumvent this time-consuming batch pre-synthesis, ligand exchange, and self-assembly steps, a powerful synthetic methodology is represented by bottom-up in situ substrate growth based on wet-chemical synthesis to form morphologically controlled nanostructures directly on the supporting substrates[24]. For MBNSs, the growth methods compatible with oxide substrates were demonstrated by first attaching colloidally prepared seeds to a substrate so that subsequent in situ overgrowth into anisotropic shapes[25,26]. This method still involves seed pre-synthesis, surface functionalization, and self-assembly steps. It shows that without prefixed seeds, the in situ growth of MBNSs can be difficult as it cannot take advantage of the native ability of the substrates, such as organosilica, hydridosilica, and other silicone-based polymers that have Si-H groups, to reduce metal salts[27,28]. Top-down lithography methods can be used to pattern seeds on the substrates[29,30], despite requiring a time-consuming lift-off and transfer process. The overall low production efficiency hinders the scalability of the device and translation into real settings.

The scientific community, therefore, requires more general and high throughput methods, to enable the exploitation of the interesting electromagnetic properties of morphology-elaborated nanostructures on substrates ready for applications, such as fiber optics for molecular sensing. To the best of our knowledge, there is a lack of literature on scalable in-situ nano-structuring on fiber facets with variable morphologies of Au nanostructures for remote SERS detection. This work is meant to fill this gap, reporting a method that can be expanded on every glass substrate.

We introduce a novel bottom-up in-situ growth approach for fabricating Au composition-homogeneous and morphology-tunable hierarchical nanostructures straightforwardly on the end facet of multi-mode optical fibers. This approach enables precise morphologically controlled nanostructuring of the fiber facet, using a two-phase process. First, we used a solid-state dewetting approach to nucleate Au nanoislands (NIs) uniformly distributed on the fiber facet, with an average diameter of approximately 50 nm and an inter-NIs distance of about 30 nm. In the second phase, we used a colloidal wet-chemical synthesis method to promote homogeneous nucleation and growth of additional Au over the surface of preexisting NI seeds. This process was facilitated by the rapid reduction of $Au^{3+}$ ions using hydroxylamine ($NH_2OH·HCl$) in the presence of varying concentrations of HEPES[31,32]. The wet-chemical procedure can be repeated twice; in the first stage, multi-dot-decorated Au NIs (MDot-NIs) are produced, whereby the density of coverage can be adjusted according to the reagent concentration. Whereas, multi-arm-decorated Au NIs (MArm-NIs) develop when the wet chemical process is performed twice; in the latter case considerable elongation and branching are favored. The whole synthetic procedure allows for the control of the shape of the resulting complex nanostructures and the extent of dot-to-rod elongation attuned to the chemical conditions. This process prevents the formation of free particles in solution within the experimental timeframe. We demonstrate a clear shape transition from surface dot-rich nanostructures to branched nanostructures, promoting a shift from a thermodynamic to a kinetic growth regime. Whether the chemical conditions enabled the control of the density of Au over-deposition on pre-existing NI seeds, the number of wet-chemical synthetic steps triggered the shape transition. The two-step wet-chemical synthetic procedure allowed the synthesis of composition-homogeneous variable morphology hierarchical nanostructures (from MDot-NIs to MArm-NIs), henceforth generally referred to as hierarchical nanoislands (HNIs), densely, uniformly, and stably distributed over the fiber facet. Moreover, multiple fibers can be simultaneously exposed to the same in-situ chemical conditions, making this process scalable, robust, and reproducible. The optical fibers engineered with HNIs exhibit superior SERS enhancement outperforming pristine only-NIs-engineered fibers; this improvement is evident in both direct-facet and through-fiber excitation schemes when the surface is functionalized with BT molecules. Meanwhile, the limit of detection (LOD) experiment conducted on an aqueous solution of R6G demonstrates that the optical fibers engineered with HNIs reduce the LOD by one order of magnitude compared to only-NI-engineered fibers in an instant and through-fiber detection scheme. Among all the developed HNIs, the MDot-NIs-engineered fiber devices achieved the lowest LOD of $10^{-7}$ M. We believe our method provides opportunities to control nanoscale surface features on simply shaped nanoparticles, enabling the fabrication of highly sensitive SERS structures at the fiber tip in a cost-effective, scalable, and efficient manner.

## *Results*

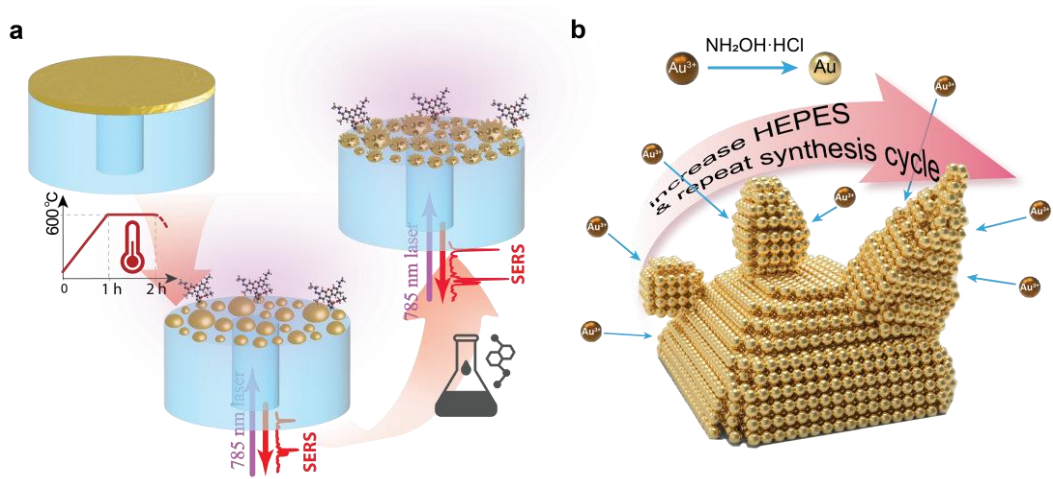

***Figure 1.*** *The sketch illustrates the method to fabricate HNIs on the fiber facets. (a) A solid-state dewetting technique is at first used to convert a thin layer of a gold film deposited across the fiber facet into hemispherical NIs that subsequently serve as seeds over which additional gold nanostructures are deposited via a HEPES-driven wet-chemical approach. (b) By adjusting the concentration of HEPES and the number of wet-chemical steps, the morphology of Au nanostructures achieved across the surface of NIs can be changed from dot- to arm-shaped.*

The overall fabrication process is illustrated in **Fig. 1**. The two-phase process combines a solid-state dewetting stage with multiple wet-chemical colloidal steps. The first solid-state high-temperature (600°C for 1 hour) dewetting stage[14] allows generating a uniform distribution of hemispherical Au NIs starting from a 5 nm thin film e-beam-evaporated onto the facet of core/cladding fibers (**Fig. 1a**). Thereafter, NIs distributed over the fiber tip serve as seeds to self-catalyze the over-deposition of additional gold in the shape of dots through a HEPES-driven wet-chemical synthesis process. The density of gold over-deposition above each NI can be adjusted according to the wet-chemical conditions. Additionally, the elongation and branching of the lately formed dot-shaped gold nanodomains can further be promoted by repeating the wet-chemical growth process (**Fig. 1b**)

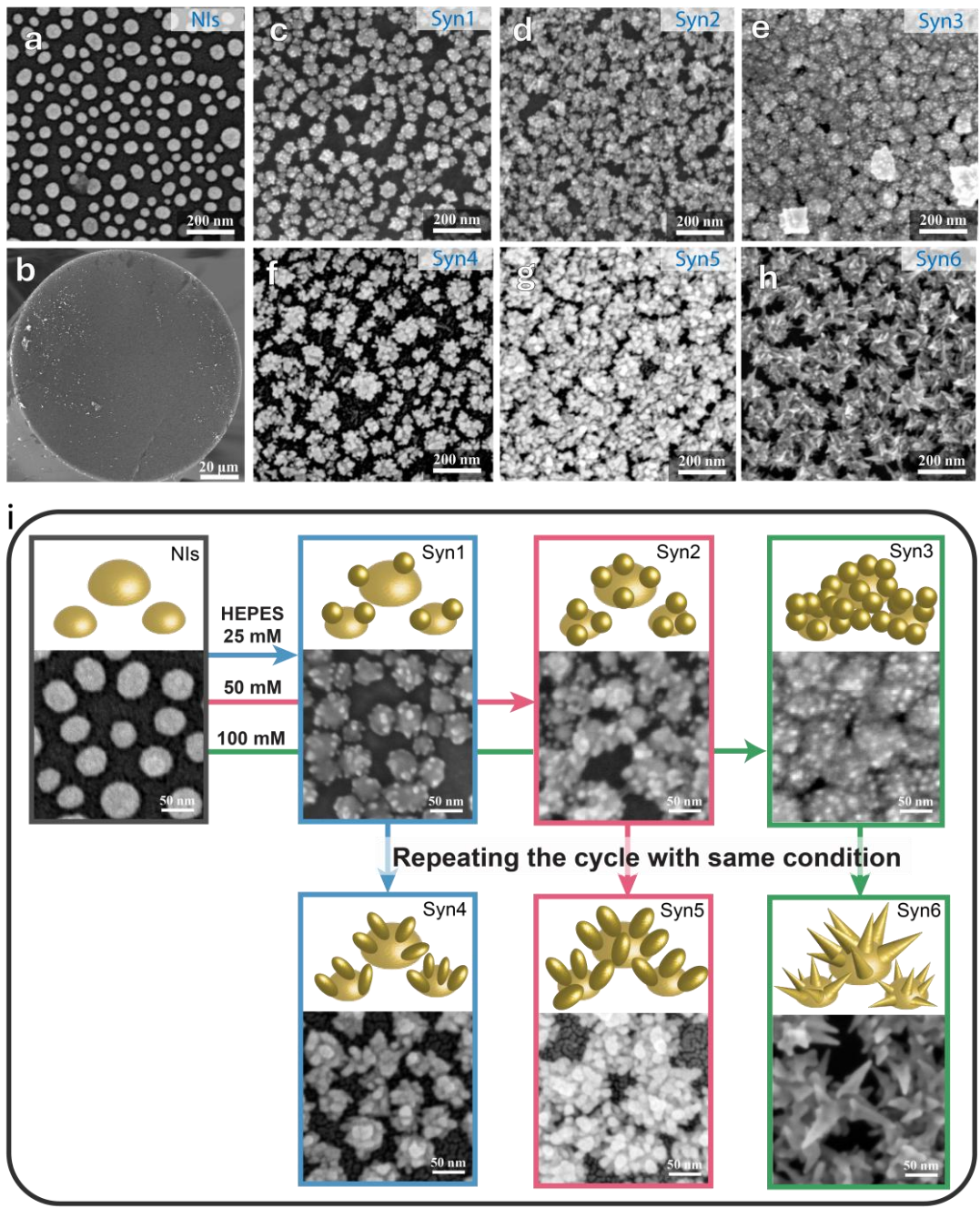

***Figure 2.*** Gallery of SEM overviews of HNIs obtained by variable chemical-driven synthesis protocols. (a) The morphology of pristine solid-state dewetted NIs as seeds prefixed on the fiber. (b) The representative fiber facet of HNIs exhibits a uniform surface status across the entire facet after synthesis. (c-e) SEM morphologies of samples (labeled as Syn1-, Syn2-, and Syn3-HNIs) were obtained using a single wet-chemical synthesis protocol with HEPES concentrations of 25 mM, 50 mM, and 100 mM in an 8 mL reaction media, respectively. (f-h) SEM morphologies of samples (labeled as Syn4-, Syn5-, and Syn6-MBNIs) were obtained using two sequential wet-chemical synthesis protocols with HEPES concentrations of 25 mM, 50 mM, and 100 mM in an 8 mL reaction media, respectively. The other synthetic conditions were unchanged. (i) Sketch illustrations and respective high-magnification SEM images depicting the morphological evolution of Au hierarchical nanostructures.

All the fabrication results are presented in **Fig. 2**; The pristine NIs seeds are shown in **Fig. 2a**. SEM images illustrating the overview of the fiber facet (**Fig. 2b**) and morphological evolution of the resulting Au nanostructures developed over the optical fiber tips are shown in **Fig. 2c-h** (large view SEM images are provided in **Fig. S1-2** for all HNIs). In detail: the nucleated NIs, with an average diameter of 50 nm and an inter-NI spacing of 30 nm, uniformly cover the entire fiber facet (**Fig. 2a**); these will serve as seeds to facilitate subsequent material nucleation via seeded growth. The tip of the NI-decorated optical fiber is immersed within an 8 mL mixture of $NH_2OH \cdot HCl$ (0.625 mM) and HEPES (25, 50, or 100 mM) for three separate experiments. In each case, 1 mL of $HAuCl_4 \cdot 3H_2O$ (1.75 mM) aqueous solution is added dropwise to the reaction media, at room temperature, while stirring. The pH is set to 7.4, whereas the HEPES concentration in the solution is changed. Generally, in these conditions, HEPES serves as a shape-directing agent and $NH_2OH \cdot HCl$ as a reducing agent of gold ions for the development of dot-like Au nanodomains. It is accepted that HEPES buffer is thermodynamically capable of working as both a reducing agent (due to cation free radical on tertiary amine of piperazine group)[32–34] as well as a shape-directing agent with its sulfonic acid group. Considering the established rate of Au addition, the time of growth is approximately 20 min (see experimental details). Under conditions of very low HEPES concentration 25 mM spherical gold dots nucleate onto preformed NIs resulting in the formation of MDot-NIs (**Fig. 2c**). By doubling the concentration to 50 mM, while always keeping constant the amount of gold, the number of dot-like domains decorating each NI increases (**Fig. 2d**); this trend continues, reaching a densely packed dot deposition at the highest HEPES concentration of 100 mM (**Fig. 2e**).

For the wet-chemical experiments just described a thermodynamic regime of growth prevails, mostly forming dot-shaped nanodomains reasonably nucleating over lattice defects exposed by the original NIs. At the very low time of growth of a wet-chemical step (20 min), the reducing ability is yielded by $NH_2OH·HCl$ and HEPES, the latter being weakly reducing at these rather dilute concentrations[35]. Nanodots nucleate and grow in such a chemical environment.

Multiple cycles of wet-chemical Au deposition can be conducted to facilitate the anisotropic development of dot domains according to the HEPES concentration. Indeed, by repeating the Au reduction in solution over fiber tips engineered by MDot-NIs, elongation and branching of dot domains occur. Under conditions of 25 mM HEPES concentration, spherical Au dots evolve into low-aspect-ratio nanorods (**Fig. 2f**) that become more elongated by doubling the HEPES concentration to 50 mM (**Fig. 2g**). With an even higher HEPES concentration of 100 mM, longer and branched arms develop generating MArm-NIs (**Fig. 2h**). This kinetic process outcome can be explained by an enhanced reducing capacity of the wet-chemical environment despite the diluted concentrations of HEPES; it can be rationalized on these bases. (i) Seed-assisted reduction: small dots earlier developed represent favorable sites of nucleation on which supplemental Au deposition occurs rather than forming new domains from scratch[33]. (ii) Surface-catalyzed reduction of $Au^{3+}$: the presence of small dots strengthens the reducing ability of $NH_2OH·HCl$[30,31,33]; the greater the number of dots, the faster the reduction. This latter kinetic effect prevents the formation of in-solution free particles because of a reductant-assisted autocatalytic deposition of Au triggered by pre-existing Au domains. In analogy with the electroless Au plating deposition, Au growth continues until the catalytic substrate is removed from the solution or the gold ion concentration is depleted. This autocatalytic Au over deposition is significantly enhanced upon highly-dot-decorated NIs. (iii) Shape-directing effect of HEPES molecules: these passivate and stabilize Au surfaces driving the formation of anisotropic and branched nanostructures as they generate templating structures with a long-range order[35]. All this evidence helps to explain the kinetic regime in the fast experimental time scale adopted and rules out possible Ostwald ripening events. A thermodynamically driven growth regime would induce a redistribution of Au atoms within the pre-existing surfaces, resulting in a smoother round-edge morphology because of atom migration from energetically unfavorable sites (such as sharp edges or tips) to decrease the surface average curvature. This process would be slower compared to the autocatalytic growth[34]. By replicating these procedures several times, more densely packed nanostructures are grown, and the Au overgrowth reduces the original NI inter-distances. It is worth noting that nucleation of Au nanostructures does not occur in the absence of dewetted NI seeds preexisting on the fiber tip (Supplementary **Fig. S3**). **Figure 2i** sketches the above-described process of Au deposition over NIs along with the high-magnification SEM images of the detailed structures. To assist the reader in referencing each sample, from this point onward, the developed structures reported in **Fig. 2c-h** will be referred to as Syn1-, Syn2-, Syn3-, Syn4-, Syn5-, and Syn6-HNIs, respectively.

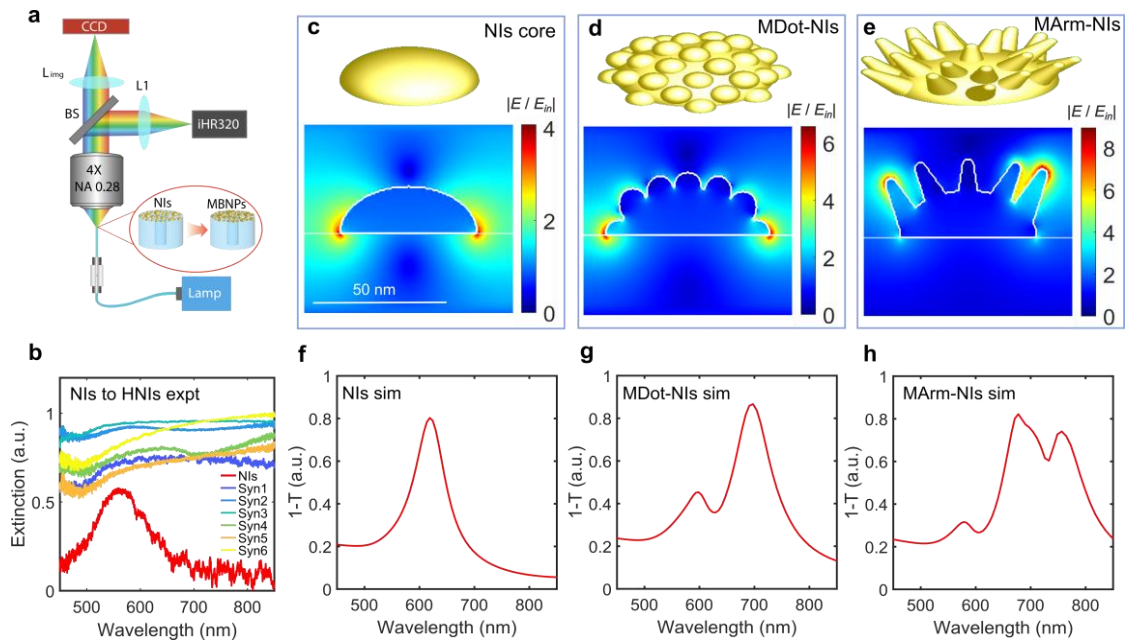

*Figure 3.* The electromagnetic response for NIs and HNIs. (a) A schematic illustration of the optical setup used for extinction measurements (BS – 50:50 beam splitter, iHR320 – Horiba spectrometer, CDD – charge coupling device). (b) Extinction spectra of fibers with facets decorated by

NIs (red) and HNIs (gradient from blue to yellow). (c-e) Simulation models for NIs, MDot-NIs, and MArm-NIs are arranged in a square periodic configuration. The core NIs have a diameter of 50 nm and a height of 15 nm, with a particle lattice periodicity of 76 nm. MDot-NIs' and MArm-NIs' models are derived from the NI model by adding 10 nm diameter gold nanospheres and nanocones (56 nm long, 20 nm wide at the base) embedded in the NIs core, respectively, across their surfaces. The field enhancements are obtained by normal incidence and 785 nm plane wave excitation. (f-h) Spectral responses of square periodic NIs, MDot-NIs, and MArm-NIs, are represented using the coefficient (1 - T), where T is the transmittance of the system at normal incidence.

To further understand the electromagnetic response of the obtained geometry-complex nanostructures, we performed extinction measurements using the configuration illustrated in **Fig. 3a**. White halogen light was guided to the plasmonic fiber facet through a fiber patch cord. The transmitted light from the fiber tip was collected by an Olympus objective (4X, NA = 0.28) and subsequently directed into a spectrometer for analysis (details in Methods). The extinction results are shown in **Fig. 3b**. The resonance transitions from single resonance peaks in NIs to broader, flattened, and increased extinction profiles in HNIs-engineered fibers. We also conducted a statistical morphological analysis to classify the obtained HNI structures. This analysis aimed to extract representative geometric parameters for constructing a simplified electromagnetic model allowing for a better understanding of the nanostructures (details provided in **Fig. S4**). Assuming normal distributions, the fitting results show that the average diameters increase from approximately 50 nm for NIs to 60 nm for Syn1-HNIs sample and the diameter distribution broadens following the overgrowth of branches. For spike structures, we measured perpendicular spikes to maximize statistical accuracy, finding an average spike length of 56 nm and a bottom width of 20 nm (detailed results in **Fig. S5**). Using these parameters, we developed a simplified numerical model consisting of a periodic square array of gold hemispheres for NIs (diameter of 50 nm, height of 15 nm and periodicity of 76 nm, parameter details are outlined in the Methods section). By decorating the surface of NIs with nanospheres (10 nm in diameter) and nanocones, we constructed models for MDot-NIs and MArm-NIs. These models enable isolation and analysis of the effects of closely packed HNIs while avoiding computationally intensive calculations. The 3D models and near-field cross-sectional distributions for NIs and HNIs in one unit cell are shown in **Fig. 3c-e**. The HNIs exhibit higher field enhancement at 785 excitations compared to NIs, with MArm-NIs showing the greatest field enhancement at the branch tip regions. The corresponding spectral responses (expressed as 1 – T, where T is the transmittance of the system at normal incidence) are presented in **Fig. 3f-h**. These results demonstrate a transition from a single resonance peak in NIs to multiple resonances in HNIs. The final optical properties of HNIs are thus found anisotropic and dependent on the morphological feature over structuring the surface of original NIs. In simulations, all complex nanostructures are uniform in size. However, in the experimental setup, individual nanostructures exhibit slight variations in size, number of dots/branches, and random arrangement. Since the HNIs are densely distributed with very small inter-distances (d << λ), the near-field electromagnetic coupling propagates across the surface in all directions. This behavior resembles a network of interacting dipole arrays[35], though with greater randomness. Consequently, the overlapping resonances of numerous HNI arrays in all directions can result in the flattened extinction spectrum observed in **Fig. 3b**.

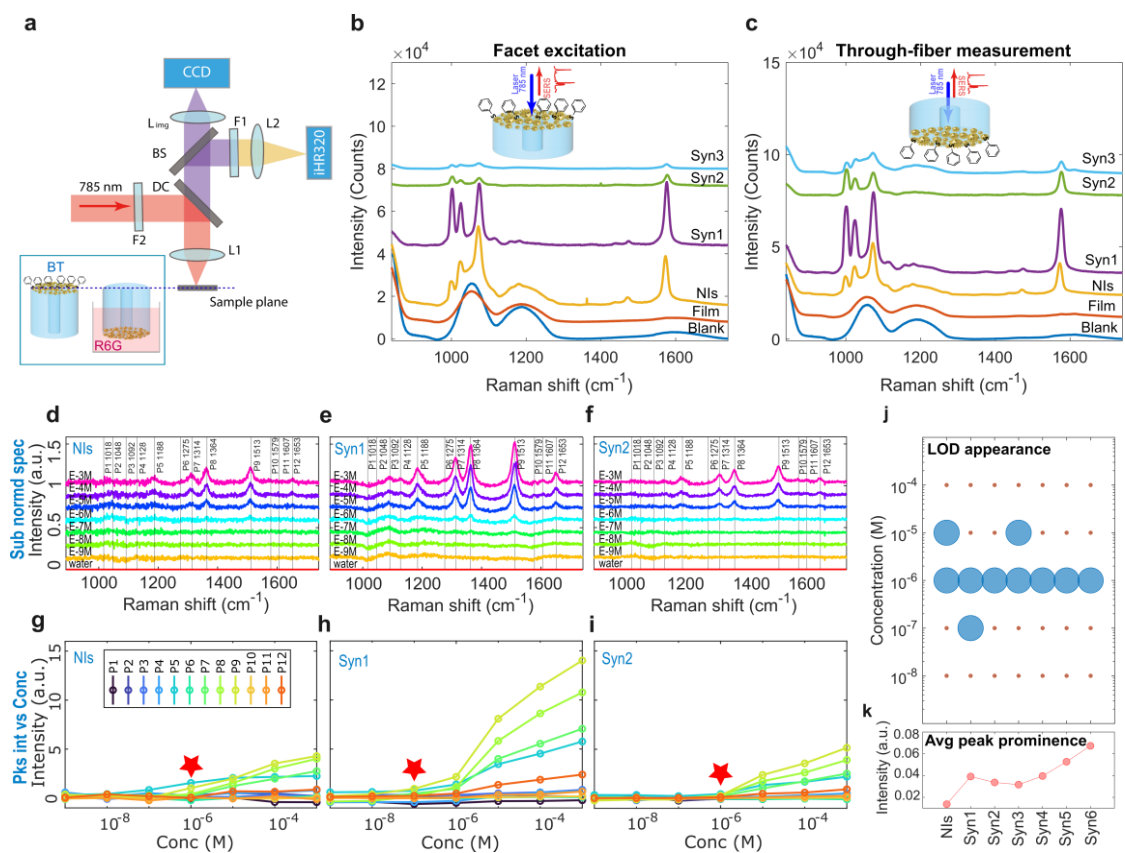

*Figure 4.* SERS characterizations for HNI fibers. (a) The schematic illustration for the optical configuration for SERS measurement; the fibers can be configured with direct facet excitation and through-fiber excitation (DC-long pass dichroic, BS-beam splitter, F1-notch filter, iHR320-spectrometer and CCD -charged coupling device). (b, c) The facet excitation and through-fiber excitation measurement with BT molecules functionalized on the fiber facet. All the fibers underwent the same functionalization procedures using the BT molecule. (d-f) The selected R6G LOD SERS spectra of fibers functionalized with NIs and Syn1/Syn2-HNIs; complete spectral sets of the reference fibers (blank and film-covered), and synthesized fibers are provided in **Fig. S6-7**; the spectra were normalized to silica peak at 1055 cm$^{-1}$, and the silica background was subtracted (the silica background was taken as the normalized spectra measured in water). The spectra sets have been vertically offset for clarity. The vertical lines mark the 12 R6G peak positions. (g-i) The peak intensities (integral of peak areas) against the concentrations, and the red stars mark the LOD for each type of fiber. (j) A summary of LOD

appearance for NIs and all the HNIs fibers; the LOD is marked as the lowest concentration when any of the R6G signature peaks of 1314, 1364, and 1513 cm$^{-1}$ become observable in the subtracted spectra. (k) The average peak prominence values for NIs and all HNIs-decorated fibers; each data point is based on an average of two measurements per condition.

The SERS performance of the fabricated probes was initially benchmarked by benzenethiol (BT) molecules functionalized on the gold surface aiming at evaluating the different performance of fibers hosting different types of nanostructures (blank, NIs, and Syn1-/Syn2-/Syn3-HNIs). Then, direct-facet excitation and through-fiber excitation of the probes were tested (optical setup in **Fig. 4a**). The results in **Fig. 4b-c** show the Syn1-HNI fiber can generate most of the SERS signal under both excitation conditions. An increase in the density of dots over NIs correlates with a decrease in signals for both Syn2- and Syn3-HNIs. This outcome suggests that the Syn1-HNIs are more effective as SERS-active nanostructures when integrated over the fiber tip. The more separated nanostructures in the Syn1-HNIs case are more efficient in generating hotspots than the more connected or continuous film-like structures in the Syn2 and Syn3 cases. Studies also have shown that dense film assemblies of Au nanostars do not necessarily create more efficient hotspots or increase the SERS enhancement[20,36,37].

Most application scenarios for SERS-active fiber probes involve remotely detecting specific chemicals in surrounding media; LOD experiments can provide insights into the performance of the fiber-optic probes. Thus, a set of fibers was prepared with various facet functionalizations: blank, film-covered, NIs, Syn1- to Syn6-HNIs. These were tested using the chemically stable analyte R6G in solution. A range of R6G aqueous solutions with concentrations from 10$^{-9}$ to 10$^{-3}$ M was prepared, and through-fiber SERS measurements were conducted using a custom-built Raman microscope (details provided in the Methods section). For each fiber, measurements started with the lowest concentration solution and progressed incrementally to the highest concentration, using the same single fiber for each concentration step. Each spectrum was recorded immediately after fully immersing the fiber tip in the aqueous solution, which remained submerged throughout the exposure period. The representative measured spectra at different concentrations are reported in **Fig. 4d-f**. These spectra were obtained by first normalizing to the silica peak at 1055 cm$^{-1}$ and then subtracting the silica background; the probe's silica background for subtraction was measured in water. The blank and film-covered fibers showed minimal detection of R6G peaks across all tested concentrations (see more details in **Fig. S6**). In contrast, the NI and all synthesized HNI fibers exhibited sensitive SERS detection at low concentrations. For a detailed comparison, we integrated the areas of 12 specific peaks (1018, 1048, 1092, 1128, 1188, 1275, 1314, 1364, 1513, 1579, 1607, and 1653 cm$^{-1}$) as a measure of peak intensity and plotted this against the concentrations. The selective results for NIs, Syn1- and Syn2-HNI fibers are reported in **Fig. 4g-i** (more detailed results in **Fig. S7**). The data shows that Syn1-HNI fibers had the lowest LOD at 10$^{-7}$ M. Notably, the peaks' prominence varied with different HNI morphologies, even when the LOD remained the same. We analyzed a statistical dataset comprising 15 fibers (including NIs and different synthesis parameters of HNI fibers), with an average of 2 fibers examined for each synthesis condition. The results in **Fig. 4j** illustrate the LOD appearance for NIs and all six synthesis conditions of HNI fibers. For the NI fibers, the LOD ranges from 10$^{-6}$ to 10$^{-5}$ M, whereas, the LOD for Syn1-HNI fibers falls within 10$^{-7}$ to 10$^{-6}$ M. Most of the other synthesis conditions demonstrate a LOD of 10$^{-6}$ M, except for Syn3-HNI fibers, which show a LOD between 10$^{-6}$ to 10$^{-5}$ M. Furthermore, the peak prominence from the statistical analysis is summarized in **Fig. 4k**. It shows that all the synthesized HNI fibers exhibit significantly higher peak prominence compared to pristine NI fibers. The peak prominence for each spectrum set of R6G LOD is determined by integrating the areas under all 12 peak intensities relative to concentration, as represented in **Fig. 4g-i**. The average values for fibers with the same synthesis conditions were used to depict the peak prominence intensity in **Fig. 4k.**

*Discussions and conclusions*

Our two-phase in situ growth method, which combines solid-state dewetting with wet-chemical synthesis, the latter under seed-mediated and surfactant-assisted growth control, enables the scalable fabrication of HNIs directly on optical fiber facets. This two-phase approach effectively addresses the limitations of traditional techniques, such as ESA and top-down lithography. By leveraging Au surface-catalyzed reduction mediated by hydroxylamine and HEPES, we achieved tunable HNI morphologies ranging from MDot- to MArm-NIs, with dense and uniform coverage. The extent of elongation and branching in the dot-to-arm shape transition can be controlled. The cross-section for the excitation of these complex multifaceted nanostructures increases compared to the original NIs[19]. This method is scalable, allowing for the simultaneous processing of multiple fibers under identical conditions, and it can be adapted for use with different glass substrates. The HNI-functionalized fibers demonstrated superior SERS performance compared to pristine NI fibers as well as higher surface area. Notably, the multi-dot-shaped Syn1-HNI fiber probes achieved the lowest LOD of $10^{-7}$ M for R6G. All types of HNI fibers generally show better signal prominence compared to pristine NI fibers (**Fig. 4k**), indicating increased field enhancement for HNIs. However, it is worth noting that while field enhancement may improve signal generation, it does not necessarily lead to a lower LOD. Several factors can affect the LOD appearance when resolving molecular signatures through fibers[17]. In our case, we should consider three main factors: (1) the ability to generate the SERS signal, (2) the ability to collect the SERS signal, and (3) the relative strength of the collected SERS signal compared to the silica background.

For different morphologies, the increased field enhancement leads to a stronger SERS signal, with the signal predominantly generated at the hot spots where the electromagnetic field is most concentrated. As shown in **Fig.3c-e**, hot spots for NIs and MDot-NIs are primarily located near the interface between the substrate and nanostructures. In contrast, MArm-NIs exhibit hot spots at the tips of the branches, elevated above the substrate.

This morphological difference influences the collection of the SERS signal through the fiber. Although NIs and MDot-NIs generate weaker signals, a larger proportion of the SERS signal can be coupled back through the device. Furthermore, as the number of dots/branches (higher density of Au deposition over NIs) increases and their size grows around the NI core, more material accumulates on the fiber surface, and the structures become more interconnected. This creates a layer that reflects more silica background. As a result, the relative strength of the molecular signals diminishes, obscuring signatures at low concentrations. Ultimately, the interplay of these factors determines the probe's performance.

In our study, the MDot-NIs of Syn1-HNI fibers, characterized by their well-defined dot-shaped protrusions and well-separated distribution, demonstrated the best LOD compared to other morphologies. Future optimizations of this approach could target several key areas: first, increasing the inter-particle distance between NIs to minimize the interconnection of branches. The solid-state dewetting technique provides significant tunability by modifying deposition parameters, such as initial film thickness and deposition rate[17,38]. Second, fine-tuning the wet-chemical synthesis to achieve shorter and sharper tips throughout the NI core will enhance field concentration and improve signal collection. Additionally, integrating these probes with other sensing technologies, such as fluorescence or electrochemical sensors, could expand their potential applications.

In conclusion, this study presents a breakthrough in the fabrication of SERS-active fiber-optic probes using an innovative, scalable, bottom-up in-situ substrate growth method. By controlling the complex geometry of Au composition-homogeneous hierarchical nanostructures through the selective growth of dot- or arm-shaped outgrowths over NIs, we achieve significant improvements in SERS sensitivity. Lower detection limits and enhanced signal-to-noise ratios demonstrate this. This approach enables the creation of high-performance optical sensors that can be seamlessly transformed into tapered fibers[14,17], enhancing the efficiency of holographic SERS endoscopic imaging[15]. Moreover, it shows great potential for practical applications, including in vivo biological monitoring and real-time environmental sensing.

## *Methods*

### **Fabricating HNIs on the fiber facet**

The standard multimode silica optical fibers (Thorlabs, FG050LGA, NA=0.22, Low-OH, Ø50 μm Core) were used for flat fibers fabrication. As a preparation step, all the fibers were put in an acetone bath for 30 min to remove the acrylate jacket. The fibers were cut with a fixed length of 7.5 cm. Then, a manual fiber cleaver (Thorlabs, XL411) was used to cut one side of the fiber to obtain a smooth top flat fiber facet. Then fibers were fixed on a batch mount in an e-beam evaporator (Thermionics laboratory, inc. e-GunTM), with their top surface normally aligned to the gold source in the crucible. A 5 nm thick gold film was evaporated on the fiber facet; the evaporation rate was kept at 0.2 Å·s$^{-1}$, with chamber pressure $< 6 \times 10^{-6}$ mbar. When the evaporation procedure finished, the fibers were detached from the mount and arranged in a ceramic bowl without any adhesive for thermal annealing in a muffle furnace (Nabertherm B180). The furnace was set to gradually ramp up from room temperature (RT) to 600 °C with a rate of 10 °C min$^{-1}$ and held at 600 °C for 1h, then allowed to cool ambiently to RT. After the NIs had been fabricated, we started with the wet chemical treatment. We selected a surfactant-assisted methodology[39] where HEPES acts as a shape-directing agent, $NH_2OH·HCl$ operates as a reducing agent for $Au^{3+}$ to $Au^0$ in the presence of Au nanodomain[40] and previously deposited Au NIs as catalytic seeding surface. A proper volume (200 to 800 μL) of HEPES (1M, with pH adjusted at 7.4 with NaOH) and 50 μL of $NH_2OH·HCl$ (100 mM) water solution was dispersed in 7.8 to 7.2 mL ultrapure water, respectively, in order to obtain 8 mL as the final volume in each reaction media. The resulting concentrations in the reaction media for HEPES are 25, 50, and 100 mM, and for $NH_2OH·HCl$ is 0.625 mM. NIs-functionalized fibers were then immersed in the solution. Then, 35 μL of $HAuCl_4·3H_2O$ (50 mM) solution was dispersed in 950 μL of water, loaded in 3 mL syringes, and slowly cast at room temperature in the gently stirred reaction media with a syringe pump (parameters: diameter 6 mm, rate 50 µL·min$^{-1}$). After the injection, the fiber is rinsed three times by dipping it in a flask with clean $H_2O$ for 20 sec. The setup allows the contemporary processing of several fibers during the same process. Moreover, the same fiber can be exposed to multiple injection processes. The HEPES (4-(2-hydroxyethyl)-1-piperazineethanesulfonic acid, $C_8H_{18}N_2O_4S$), NaOH, $NH_2OH·HCl$, and $HAuCl_4·3H_2O$ were purchased from Sigma Aldrich.

**The extinction measurements for NIs and HNIs fibers**

The extinction spectra were measured with a transmission configuration using a home-built microscope. Briefly, a halogen lamp with an SMA connector was used to excite the fiber through a fiber patch-cord connection, which delivered the broadband light through the NIs and HNPs functionalized fiber facet. The transmitted light was collected with a 4X objective (Olympus XLFluor 4x/340, NA = 0.28), and then directed to a spectrometer (Horiba iHR320) with an achromatic doublet (Ø25.4 mm, f = 100 mm). The spectrometer is equipped with a 300 l·mm$^{-1}$ grating and synapse EMCCD for the spectra acquisition. To subtract the actual extinction spectrum of NI and HNI-functionalized fibers ($Ext_{NIs}$), the transmission spectrum of NI and HNI fibers ($T_{NIs}$) is corrected by a blank fiber transmission spectrum ($T_{blank}$) with relation $Ext_{NIs} = (T_{blank} - T_{NIs})/T_{blank}$, to exclude the systematic response.

**The morphological analysis for NIs and HNIs**

SEM images were acquired with the FEI Helios Nanolab 600i Dual Beam system for the morphological analysis. The SEM image acquisitions were usually conducted after sputtering a thin layer of gold on the NI fiber' surface, while some chemically synthesized HNI fibers can have enough conductivity for SEM imaging without any sputtering. Home-developed MATLAB programs were used for NI morphological analysis. By tracing object boundaries, we obtain the NI number count and the coverage situations. We extracted the coverage rate ($C_1$ = 33.34%) and the areas of physical occupation ($S_1$), then the diameters were determined by $D_1 = $

$2 \times \sqrt{S/\pi}$, the result is shown in the histogram in **Fig. S4**. The average height *($H_1$ = 15 nm)* of the NIs can be computed by $H_1 = Thk/C_1$, where *Thk* = 5 nm is the initial film deposition thickness. For the Syn1-HNI analysis, as the HNIs are more branching each other, we only extracted the coverage rate ($C_2$ = 53%) by tracing object boundaries using gray level, then we used a manual way to extract the diameters of Syn1-HNIs and included nanoparticles as many as possible within the obtained SEM images to form statistics by ImageJ. For the MArm-NIs (Syn6-HNIs), we measured the perpendicular spikes to maximize statistical accuracy. We found that the average spike length is 56 nm, and the bottom width is 20 nm. Detailed results can be found in **Fig. S5**.

**Electromagnetic simulations.**
The geometric parameters obtained from morphological analysis, including NI average diameters ($D_1$), height ($H_1$), and coverage rate ($C_1$), were used to build up a simplified numerical model using the finite difference time domain (FDTD) method (Ansys-Lumerical), to emphasize the effect of the interparticle distances and reduce the computational cost. Thus, the NI pattern is represented with an Au droplet-like structure (to mimic the NI shape) arranged in a periodic square array, having a diameter of 50 nm and a maximum height of 15 nm. The period of the unit cell was set as 76 nm. The $C_1$ is defined as the ratio between the gold disk area and the square unit cell area. The $C_1$ and $D_1$ are known from the morphological analysis. The same square periodic arrangement was used as NIs for HNIs. As the diameters of NIs and Syn1-HNIs are around 50 nm to 60 nm, thus we use nanospheres of 10 nm diameter embedded into the NI core to form the HNIs. For MArm-NIs, the nanocones (with a length of 56 nm and a bottom width of 20 nm) were randomly integrated with NIs to mimic the branched structure. By simulating three-dimensional time-harmonic Maxwell's equations, we obtained the results in **Fig. 3**. The particle sits on an infinite dielectric substrate with a refractive index n = 1.4. Optical constants of gold were obtained from Ref.[41]. The spectral response was obtained by scanning the source's wavelength from 400 to 900 nm, while field enhancement maps were generated at 785 nm, a widely employed wavelength for Raman inspection of biological samples.

**SERS characterizations on the fibers**

To prepare the fibers for optical characterizations, the unstructured side of fabricated fibers was connected to metallic ferrules with a diameter of 1.25 mm and underwent a manual polishing process. For the optical measurements, a home-built Raman microscope is used for characterization; further details can be found in our previous publication[17]. Briefly, the linear polarized free space laser of 785 nm continuous wavelength was coupled into a meter-long fiber patch cord to launch the laser into the excitation path of the Raman microscope, and the resulting excitation laser delivered to the sample was scrambled into a speckle pattern. The collimated laser beam filled the back aperture of the focus lens L1 (aspheric, Ø25.0 mm, f = 20 mm), resulting in a light spot of 50 μm in diameter, which matches the fiber core size. For the facet excitation, the fibers were configured with a proximal end facing the focus lens with a power of 6 mW. In the through-fiber excitation, the fibers were configured with the distal end facing the focus lens with a power of 10 mW; the laser was injected over the full angular acceptance of the fiber (NA = 0.22) to recruit most of the propagating modes. The Raman signals were then separated from the pump laser using a dichroic mirror (DC: Semrock, LPD02-785RU-25) and a long-pass razor-edge filter (F1: Semrock, LP02-785RU-25). Then, the signal was routed to a spectrometer (Horiba iHR320). The Raman measurements were performed with a slit at 200 μm and a 600 $l \cdot mm^{-1}$ (blaze 750 nm) grating. Spectra were recorded on a SYNAPSE CCD cooled to -50 °C. All the raw spectra were treated with baseline correction (ALS). In the through-fiber SERS measurements. The spectra acquisition time is 60 sec for all measurements. For the BT molecule functionalized fiber measurements, all reference fibers were immersed in a 6 mM methanol solution of BT molecule for 6.5 h. Afterward, the fibers were rinsed by stirring them in a cup of clean methanol solution for 10 mins. This rinsing process was repeated three times, using fresh methanol solution each time, to ensure successful monolayer functionalization. Both BT (benzenethiol, $C_6H_5SH$) and R6G

(Rhodamine 6G, $C_{28}H_{31}N_2O_3Cl$) were purchased from Merck KGaA.

*Acknowledgments*

F.P. and L.C. jointly supervised and are co-last authors in this work.
M.D.V. and F.P. acknowledge funding by European Union's Horizon Europe under grant agreement 101125498, project MINING - "Multifunctional nano-bio interfaces with deep brain regions". D.Z., M.F.K., Li.C., M.D.V., and F.P. acknowledge funding from the European Union's Horizon 2020 Research and Innovation Program under Grant Agreement No. 828972. Li.C., M.D.V., and F.P. acknowledge funding from the Project "RAISE (Robotics and AI for Socio-economic Empowerment)" code ECS00000035 funded by European Union – NextGenerationEU PNRR MUR - M4C2 – Investimento 1.5 - Avviso "Ecosistemi dell'Innovazione" CUP J33C22001220001. M.D.V. and F.P. acknowledge that this project has received funding from the European Union's Horizon 2020 Research and Innovation Program under Grant Agreement No. 101016787. L.C. and R.S. acknowledge the Ministry of University and Research (Directorial Decree No. 1409 of 14-9-2022) within the project PRIN 2022 PNRR "CO2@photothermocat - Removal of air pollutants and valorization of the produced CO2: hybrid catalysis to solve two issues at single blow", P2022JXKKF_PE4 - CUP B53D23025390001 funded by the European Union - Next Generation EU. Moreover, L.C. and R.S. acknowledge the Italian Ministry of Research (MUR) in the framework of the National Recovery and Resilience Plan (NRRP) and under the complementary actions to the NRRP funded by NextGenerationEU, "Fit4MedRob" Grant (PNC0000007, B53C22006960001)

*Reference*

(1) Bergholt, M. S.; Zheng, W.; Ho, K. Y.; Teh, M.; Yeoh, K. G.; So, J. B. Y.; Shabbir, A.; Huang, Z. Fiber-Optic Raman Spectroscopy Probes Gastric Carcinogenesis in Vivo at Endoscopy. *Journal of Biophotonics* **2013**, *6* (1), 49–59. https://doi.org/10.1002/jbio.201200138.
(2) McGregor, H. C.; Short, M. A.; Lam, S.; Shaipanich, T.; Beaudoin, E.-L.; Zeng, H. Development and in Vivo Test of a Miniature Raman Probe for Early Cancer

Detection in the Peripheral Lung. *Journal of Biophotonics* **2018**, *11* (11), e201800055. https://doi.org/10.1002/jbio.201800055.

(3) Fortuni, B.; Ricci, M.; Vitale, R.; Inose, T.; Zhang, Q.; Hutchison, J. A.; Hirai, K.; Fujita, Y.; Toyouchi, S.; Krzyzowska, S.; Van Zundert, I.; Rocha, S.; Uji-i, H. SERS Endoscopy for Monitoring Intracellular Drug Dynamics. *ACS Sens.* **2023**, *8* (6), 2340–2347. https://doi.org/10.1021/acssensors.3c00394.

(4) Zhang, Q.; Inose, T.; Ricci, M.; Li, J.; Tian, Y.; Wen, H.; Toyouchi, S.; Fron, E.; Ngoc Dao, A. T.; Kasai, H.; Rocha, S.; Hirai, K.; Fortuni, B.; Uji-i, H. Gold-Photodeposited Silver Nanowire Endoscopy for Cytosolic and Nuclear pH Sensing. *ACS Appl. Nano Mater.* **2021**, *4* (9), 9886–9894. https://doi.org/10.1021/acsanm.1c02363.

(5) Zhang, H.; Zhou, X.; Li, X.; Gong, P.; Zhang, Y.; Zhao, Y. Recent Advancements of LSPR Fiber-Optic Biosensing: Combination Methods, Structure, and Prospects. *Biosensors* **2023**, *13* (3), 405. https://doi.org/10.3390/bios13030405.

(6) Liu, Y.; Lin, C.; Chen, H.; Shen, C.; Zheng, Z.; Li, M.; Xu, B.; Zhao, C.; Kang, J.; Wang, Y. A Rapid Surface-Enhanced Raman Scattering Method for the Determination of Trace Hg2+ with Tapered Optical Fiber Probe. *Microchemical Journal* **2024**, *196*, 109724. https://doi.org/10.1016/j.microc.2023.109724.

(7) Lyu, D.; Huang, Q.; Wu, X.; Nie, Y.; Yang, M. Optical Fiber Sensors for Water and Air Quality Monitoring: A Review. *OE* **2023**, *63* (3), 031004. https://doi.org/10.1117/1.OE.63.3.031004.

(8) Zheng, D.; Zhang, S.; Deng, Q.; Kang, M.; Nordlander, P.; Xu, H. Manipulating Coherent Plasmon–Exciton Interaction in a Single Silver Nanorod on Monolayer WSe$_2$. *Nano Letters* **2017**, *17* (6), 3809–3814. https://doi.org/10.1021/acs.nanolett.7b01176.

(9) Schuller, J. A.; Barnard, E. S.; Cai, W.; Jun, Y. C.; White, J. S.; Brongersma, M. L. Plasmonics for Extreme Light Concentration and Manipulation. *Nature Materials* **2010**, *9* (3), 193–204. https://doi.org/10.1038/nmat2630.

(10) Xu, H.; Bjerneld, E. J.; Käll, M.; Börjesson, L. Spectroscopy of Single Hemoglobin Molecules by Surface Enhanced Raman Scattering. *Phys. Rev. Lett.* **1999**, *83* (21), 4357–4360. https://doi.org/10.1103/PhysRevLett.83.4357.

(11) Hendriks, A. L.; Rabelink, D.; Dolci, M.; Dreverman, P.; Cano-Velázquez, M. S.; Picelli, L.; Veldhoven, R. P. J. van; Zijlstra, P.; Verhagen, E.; Fiore, A. Detecting Single Nanoparticles Using Fiber-Tip Nanophotonics. *Optica, OPTICA* **2024**, *11* (4), 512–518. https://doi.org/10.1364/OPTICA.516575.

(12) Deng, Q.; Kang, M.; Zheng, D.; Zhang, S.; Xu, H. Mimicking Plasmonic Nanolaser Emission by Selective Extraction of Electromagnetic Near-Field from Photonic Microcavity. *Nanoscale* **2018**, *10* (16), 7431–7439. https://doi.org/10.1039/C8NR00102B.

(13) Gangwar, R. K.; Pathak, A. K.; Chiavaioli, F.; Abu Bakar, M. H.; Kamil, Y. M.; Mahdi, M. A.; Singh, V. K. Optical Fiber SERS Sensors: Unveiling Advances, Challenges, and Applications in a Miniaturized Technology. *Coordination Chemistry Reviews* **2024**, *510*, 215861. https://doi.org/10.1016/j.ccr.2024.215861.

(14) Zheng, D.; Pisano, F.; Collard, L.; Balena, A.; Pisanello, M.; Spagnolo, B.; Mach-Batlle, R.; Tantussi, F.; Carbone, L.; De Angelis, F.; Valiente, M.; de la Prida, L. M.; Ciracì, C.; De Vittorio, M.; Pisanello, F. Toward Plasmonic Neural Probes: SERS Detection of Neurotransmitters through Gold-Nanoislands-Decorated Tapered Optical Fibers with Sub-10 Nm Gaps. *Advanced Materials* **2023**, *35* (11), 2200902. https://doi.org/10.1002/adma.202200902.

(15) Collard, L.; Pisano, F.; Zheng, D.; Balena, A.; Kashif, M. F.; Pisanello, M.; D'Orazio, A.; de la Prida, L. M.; Ciracì, C.; Grande, M.; De Vittorio, M.; Pisanello, F. Holographic Manipulation of Nanostructured Fiber Optics Enables Spatially-Resolved, Reconfigurable Optical Control of Plasmonic Local Field Enhancement and SERS. *Small* **2022**, *18* (23), 2200975. https://doi.org/10.1002/smll.202200975.

(16) Pisano, F.; Kashif, M. F.; Balena, A.; Pisanello, M.; De Angelis, F.; de la Prida, L. M.; Valiente, M.; D'Orazio, A.; De Vittorio, M.; Grande, M.; Pisanello, F. Plasmonics on a Neural Implant: Engineering Light–Matter Interactions on the Nonplanar Surface of Tapered Optical Fibers. *Adv. Opt. Mater.* **2022**, *10* (2), 2101649. https://doi.org/10.1002/adom.202101649.

(17) Zheng, D.; Kashif, M. F.; Piscopo, L.; Collard, L.; Ciracì, C.; De Vittorio, M.; Pisanello, F. Tunable Nanoislands Decorated Tapered Optical Fibers Reveal Concurrent Contributions in Through-Fiber SERS Detection. *ACS Photonics* **2024**, *11* (9), 3774–3783. https://doi.org/10.1021/acsphotonics.4c00912.

(18) Ngo, N. M.; Tran, H.-V.; Lee, T. R. Plasmonic Nanostars: Systematic Review of Their Synthesis and Applications. *ACS Appl. Nano Mater.* **2022**, *5* (10), 14051–14091. https://doi.org/10.1021/acsanm.2c02533.

(19) Hao, F.; Nehl, C. L.; Hafner, J. H.; Nordlander, P. Plasmon Resonances of a Gold Nanostar. *Nano Lett.* **2007**, *7* (3), 729–732. https://doi.org/10.1021/nl062969c.

(20) Solís, D. M.; Taboada, J. M.; Obelleiro, F.; Liz-Marzán, L. M.; García de Abajo, F. J. Optimization of Nanoparticle-Based SERS Substrates through Large-Scale Realistic Simulations. *ACS Photonics* **2017**, *4* (2), 329–337. https://doi.org/10.1021/acsphotonics.6b00786.

(21) Tian, Q.; Cao, S.; He, G.; Long, Y.; Zhou, X.; Zhang, J.; Xie, J.; Zhao, X. Plasmonic Au-Ag Alloy Nanostars Based High Sensitivity Surface Enhanced Raman Spectroscopy Fiber Probes. *Journal of Alloys and Compounds* **2022**, *900*, 163345. https://doi.org/10.1016/j.jallcom.2021.163345.

(22) He, G.; Han, X.; Cao, S.; Cui, K.; Tian, Q.; Zhang, J. Long Spiky Au-Ag Nanostar Based Fiber Probe for Surface Enhanced Raman Spectroscopy. *Materials* **2022**, *15* (4), 1498. https://doi.org/10.3390/ma15041498.

(23) Zhao, X.; Campbell, S.; Wallace, G. Q.; Claing, A.; Bazuin, C. G.; Masson, J.-F. Branched Au Nanoparticles on Nanofibers for Surface-Enhanced Raman Scattering Sensing of Intracellular pH and Extracellular pH Gradients. *ACS Sens.* **2020**, *5* (7), 2155–2167. https://doi.org/10.1021/acssensors.0c00784.

(24) Vinnacombe-Willson, G. A.; Conti, Y.; Stefancu, A.; Weiss, P. S.; Cortés, E.; Scarabelli, L. Direct Bottom-Up In Situ Growth: A Paradigm Shift for Studies in Wet-Chemical Synthesis of Gold Nanoparticles. *Chem. Rev.* **2023**, *123* (13), 8488–8529. https://doi.org/10.1021/acs.chemrev.2c00914.

(25) Vinnacombe-Willson, G. A.; Lee, J. K.; Chiang, N.; Scarabelli, L.; Yue, S.; Foley, R.; Frost, I.; Weiss, P. S.; Jonas, S. J. Exploring the Bottom-Up Growth of Anisotropic Gold Nanoparticles from Substrate-Bound Seeds in Microfluidic Reactors. *ACS Appl. Nano Mater.* **2023**, *6* (8), 6454–6460. https://doi.org/10.1021/acsanm.3c00440.

(26) Vinnacombe-Willson, G. A.; Chiang, N.; Scarabelli, L.; Hu, Y.; Heidenreich, L. K.; Li, X.; Gong, Y.; Inouye, D. T.; Fisher, T. S.; Weiss, P. S.; Jonas, S. J. In Situ Shape Control of Thermoplasmonic Gold Nanostars on Oxide Substrates for Hyperthermia-Mediated Cell Detachment. *ACS Cent. Sci.* **2020**, *6* (11), 2105–2116. https://doi.org/10.1021/acscentsci.0c01097.

(27) Fortuni, B.; Fujita, Y.; Ricci, M.; Inose, T.; Aubert, R.; Lu, G.; Hutchison, J. A.; Hofkens, J.; Latterini, L.; Uji-i, H. A Novel Method for in Situ Synthesis of SERS-Active Gold Nanostars on Polydimethylsiloxane Film. *Chem. Commun.* **2017**, *53* (37), 5121–5124. https://doi.org/10.1039/C7CC01776F.

(28) Vinnacombe-Willson, G. A.; Conti, Y.; Jonas, S. J.; Weiss, P. S.; Mihi, A.; Scarabelli, L. Surface Lattice Plasmon Resonances by Direct In Situ Substrate Growth of Gold Nanoparticles in Ordered Arrays. *Advanced Materials* **2022**, *34* (37), 2205330. https://doi.org/10.1002/adma.202205330.

(29) Jia, J.; Metzkow, N.; Park, S.-M.; Wu, Y. L.; Sample, A. D.; Diloknawarit, B.; Jung, I.; Odom, T. W. Spike Growth on Patterned Gold Nanoparticle Scaffolds. *Nano Lett.* **2023**, *23* (23), 11260–11265. https://doi.org/10.1021/acs.nanolett.3c03778.

(30) Shen, Y.; Liang, L.; Zhang, S.; Huang, D.; Zhang, J.; Xu, S.; Liang, C.; Xu, W. Large-Area Periodic Arrays of Gold Nanostars Derived from HEPES-, DMF-, and Ascorbic-Acid-Driven Syntheses. *Nanoscale* **2018**, *10* (4), 1622–1630. https://doi.org/10.1039/C7NR08636A.

(31) Stremsdoerfer, G.; Perrot, H.; Martin, J. R.; Cléchet, P. Autocatalytic Deposition of Gold and Palladium onto n - GaAs in Acidic Media. *J. Electrochem. Soc.* **1988**, *135* (11), 2881. https://doi.org/10.1149/1.2095453.

(32) Brown, K. R.; Natan, M. J. Hydroxylamine Seeding of Colloidal Au Nanoparticles in Solution and on Surfaces. *Langmuir* **1998**, *14* (4), 726–728. https://doi.org/10.1021/la970982u.

(33) Yang, T.-H.; Zhou, S.; Gilroy, K. D.; Figueroa-Cosme, L.; Lee, Y.-H.; Wu, J.-M.; Xia, Y. Autocatalytic Surface Reduction and Its Role in Controlling Seed-Mediated Growth of Colloidal Metal Nanocrystals. *Proceedings of the National Academy of Sciences* **2017**, *114* (52), 13619–13624. https://doi.org/10.1073/pnas.1713907114.

(34) Winkler, K.; Kaminska, A.; Wojciechowski, T.; Holyst, R.; Fialkowski, M. Gold Micro-Flowers: One-Step Fabrication of Efficient, Highly Reproducible Surface-Enhanced Raman Spectroscopy Platform. *Plasmonics* **2011**, *6* (4), 697–704. https://doi.org/10.1007/s11468-011-9253-0.

(35) Maier, S. A.; Brongersma, M. L.; Kik, P. G.; Atwater, H. A. Observation of Near-Field Coupling in Metal Nanoparticle Chains Using Far-Field Polarization Spectroscopy. *Phys. Rev. B* **2002**, *65* (19), 193408. https://doi.org/10.1103/PhysRevB.65.193408.

(36) Langer, J.; Aberasturi, D. J. de; Aizpurua, J.; Alvarez-Puebla, R. A.; Auguié, B.; Baumberg, J. J.; Bazan, G. C.; Bell, S. E. J.; Boisen, A.; Brolo, A. G.; Choo, J.; Cialla-May, D.; Deckert, V.; Fabris, L.; Faulds, K.; Abajo, F. J. G. de; Goodacre, R.; Graham, D.; Haes, A. J.; Haynes, C. L.; Huck, C.; Itoh, T.; Käll, M.; Kneipp, J.; Kotov, N. A.; Kuang, H.; Ru, E. C. L.; Lee, H. K.; Li, J.-F.; Ling, X. Y.; Maier, S. A.; Mayerhöfer, T.; Moskovits, M.; Murakoshi, K.; Nam, J.-M.; Nie, S.; Ozaki, Y.; Pastoriza-Santos, I.; Perez-Juste, J.; Popp, J.; Pucci, A.; Reich, S.; Ren, B.; Schatz, G. C.; Shegai, T.; Schlücker, S.; Tay, L.-L.; Thomas, K. G.; Tian, Z.-Q.; Duyne, R. P. V.; Vo-Dinh, T.; Wang, Y.; Willets, K. A.; Xu, C.; Xu, H.; Xu, Y.; Yamamoto, Y. S.; Zhao, B.; Liz-Marzán, L. M. Present and Future of Surface-Enhanced Raman Scattering. *ACS Nano* **2019**. https://doi.org/10.1021/acsnano.9b04224.

(37) Serrano-Montes, A. B.; Jimenez de Aberasturi, D.; Langer, J.; Giner-Casares, J. J.; Scarabelli, L.; Herrero, A.; Liz-Marzán, L. M. A General Method for Solvent Exchange of Plasmonic Nanoparticles and Self-Assembly into SERS-Active Monolayers. *Langmuir* **2015**, *31* (33), 9205–9213. https://doi.org/10.1021/acs.langmuir.5b01838.

(38) Tesler, A. B.; Chuntonov, L.; Karakouz, T.; Bendikov, T. A.; Haran, G.; Vaskevich, A.; Rubinstein, I. Tunable Localized Plasmon Transducers Prepared by Thermal Dewetting of Percolated Evaporated Gold Films. *J. Phys. Chem. C* **2011**, *115* (50), 24642–24652. https://doi.org/10.1021/jp209114j.

(39) Maiorano, G.; Rizzello, L.; Malvindi, M. A.; Shankar, S. S.; Martiradonna, L.; Falqui, A.; Cingolani, R.; Pompa, P. P. Monodispersed and Size-Controlled Multibranched Gold Nanoparticles with Nanoscale Tuning of Surface Morphology. *Nanoscale* **2011**, *3* (5), 2227. https://doi.org/10.1039/c1nr10107b.

(40) Stremsdoerfer, G.; Perrot, H.; Martin, J. R.; Cléchet, P. Autocatalytic Deposition of Gold and Palladium onto n - GaAs in Acidic Media. *J. Electrochem. Soc.* **1988**, *135* (11), 2881. https://doi.org/10.1149/1.2095453.

(41) Olmon, R. L.; Slovick, B.; Johnson, T. W.; Shelton, D.; Oh, S.-H.; Boreman, G. D.; Raschke, M. B. Optical Dielectric Function of Gold. *Phys. Rev. B* **2012**, *86* (23), 235147. https://doi.org/10.1103/PhysRevB.86.235147.

**Supporting Information for**

**Direct Nucleation of Hierarchical Nanostructures on Plasmonic Fiber**

**Optics Enables Enhanced SERS Performance**

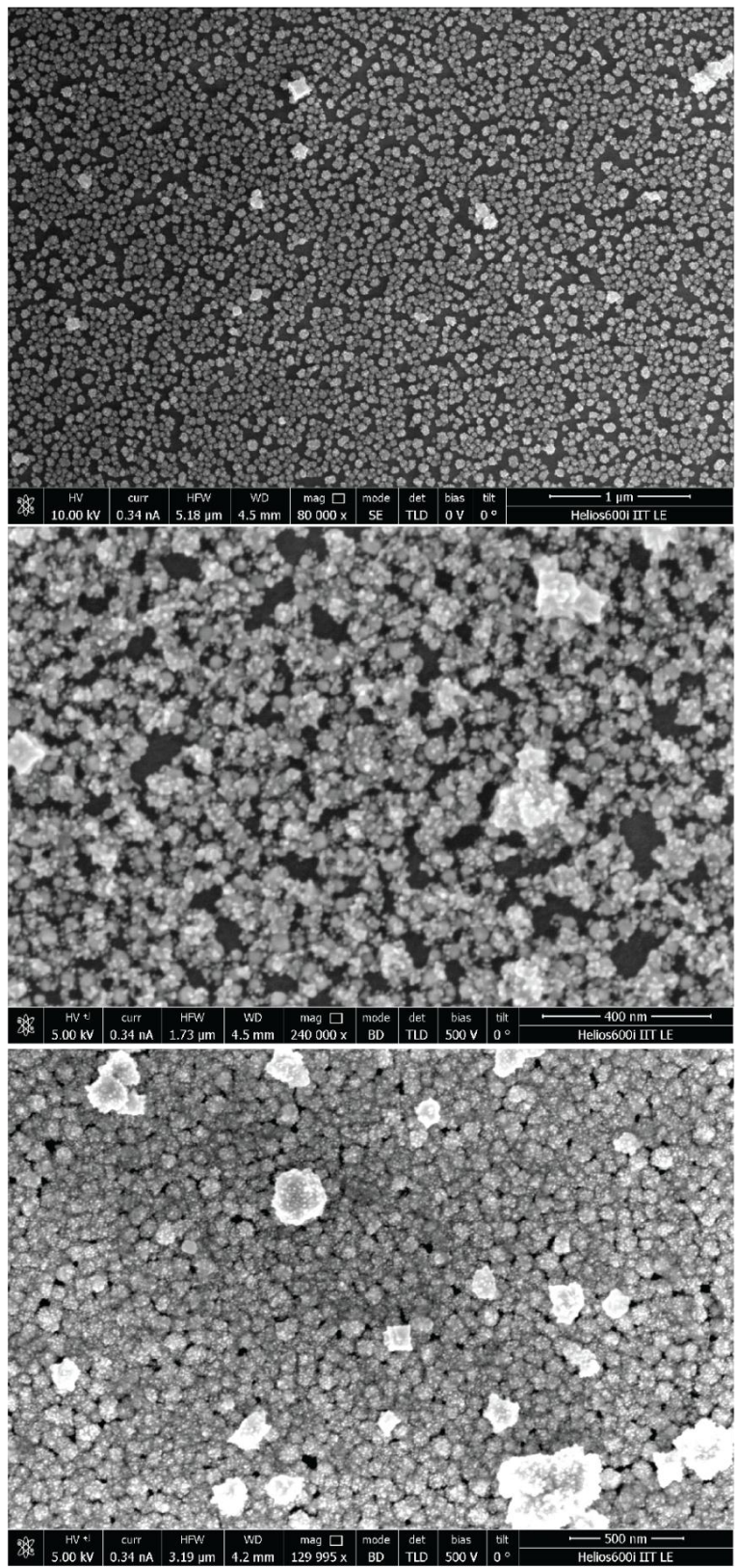

*Figure S1* SEM overview of HNIs obtained through a single wet-chemical synthesis protocol; from top to bottom are Syn1-, Syn2-, and Syn3-HNIs respectively.

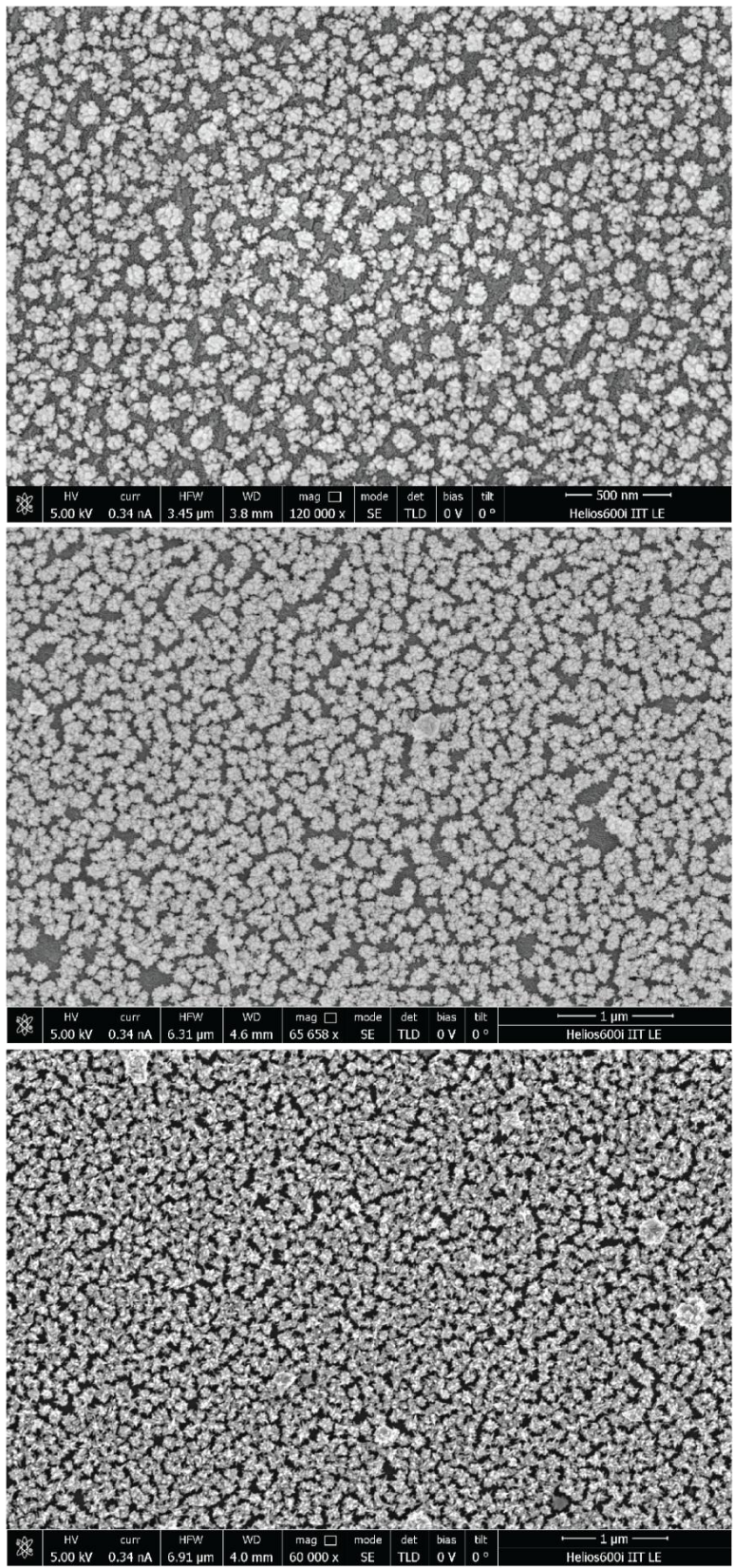

*Figure S2* SEM overview of HNIs obtained through two sequential wet-chemical synthesis protocol; from top to bottom are Syn4-, Syn5-, and Syn6-HNIs respectively.

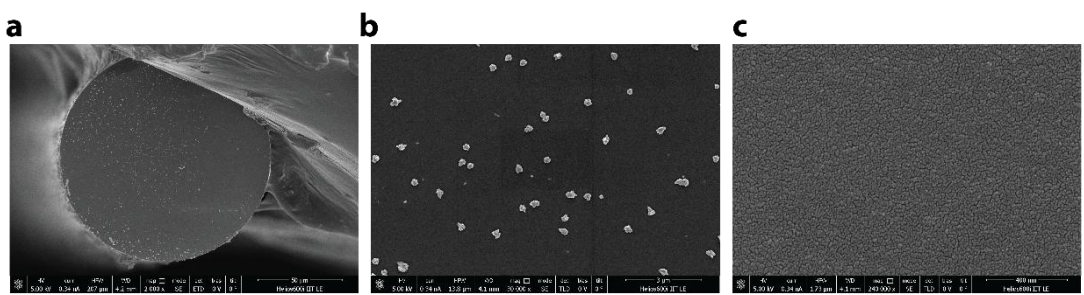

***Figure S3*** *The morphology of a blank fiber (without NI seeds) subjected to the synthesis protocol of Syn4. (a) Overview of the fiber facet. (b, c) Magnified views of the fiber facet morphology. Aside from some intermediate products adhering to the fiber facet, no evidence of on-substrate particle growth was observed. The grainy texture visible in the highly magnified images corresponds to sputtered gold, which was applied to ensure sufficient conductivity for SEM imaging.*

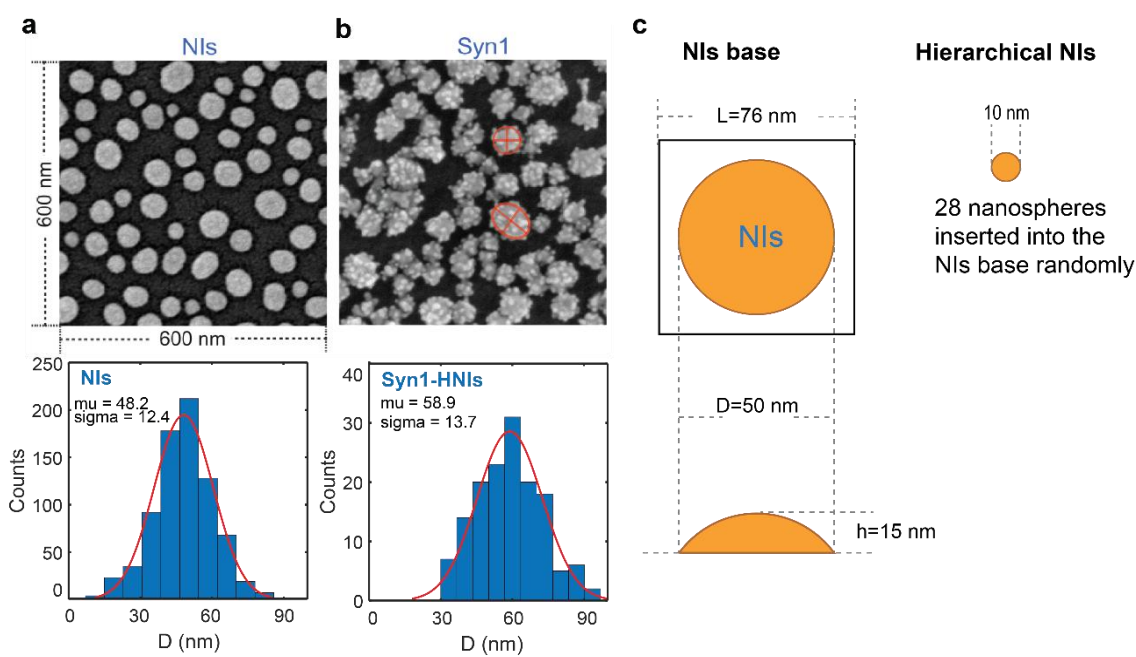

***Figure S4*** *(a) Morphology of the NIs (upper panel) and corresponding diameter distribution (lower panel). The NIs were extracted using gray-level difference analysis. (b) Morphology of the Syn1-HNIs (upper panel) and diameter analysis (lower panel). Diameters were determined by averaging measurements from all manually assessed HNIs, with two-axis measurements taken for each particle based on the profile. Two examples are highlighted with red circles and crosses in the SEM image. (c) Square periodic model for the NIs based on average diameter and coverage rate. The Syn1-HNIs model was constructed by randomly inserting 28 nanospheres into the base of the NIs.*

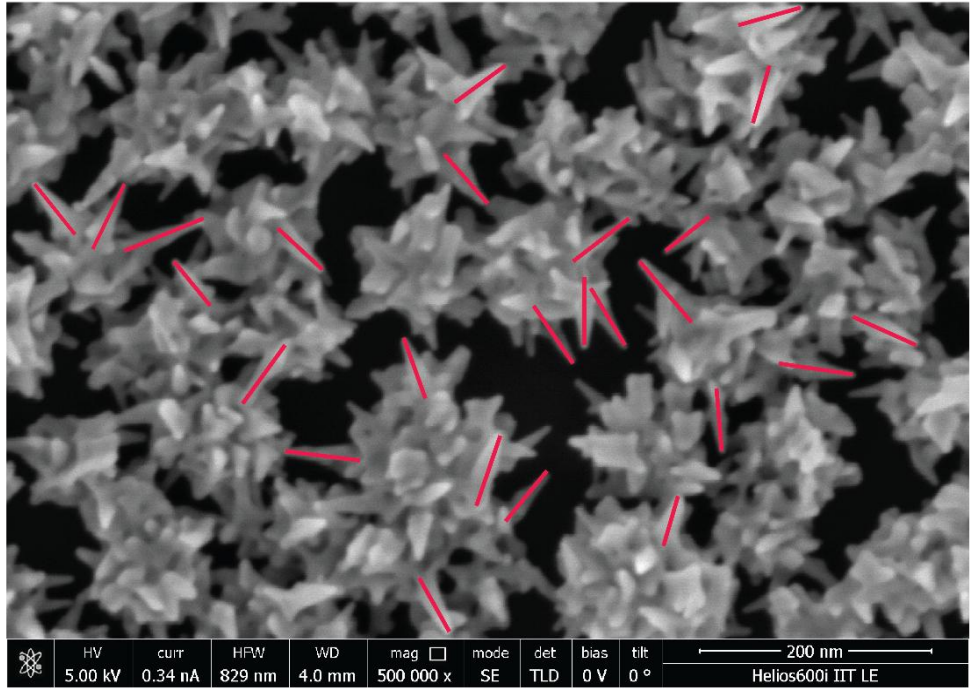

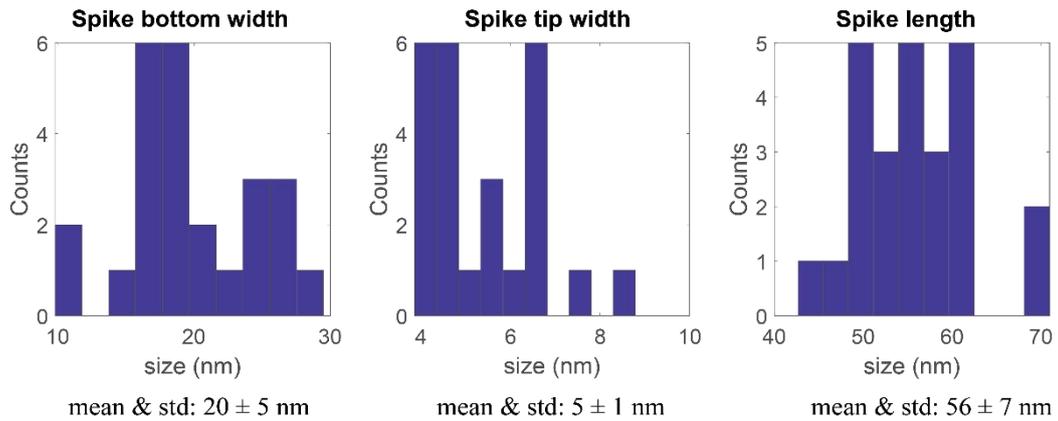

**The model of nanocone:**

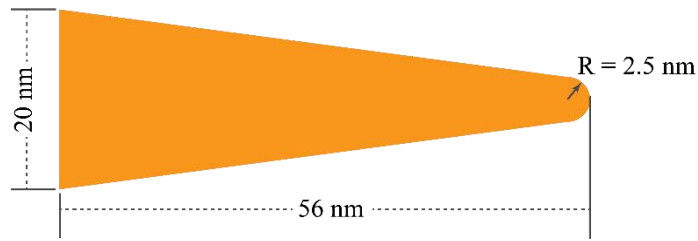

***Figure S5*** *Morphology analysis of Syn6-HNIs. Perpendicular spikes were manually measured, as indicated by the red lines in the SEM image. The bar graphs display the measured dimensions of the spikes. Using the average geometric parameters extracted from the SEM image, the spikes were modeled as nanocones for simulation, with a base length of 20 nm, a side length of 56 nm, and a tip rounding of 2.5 nm.*

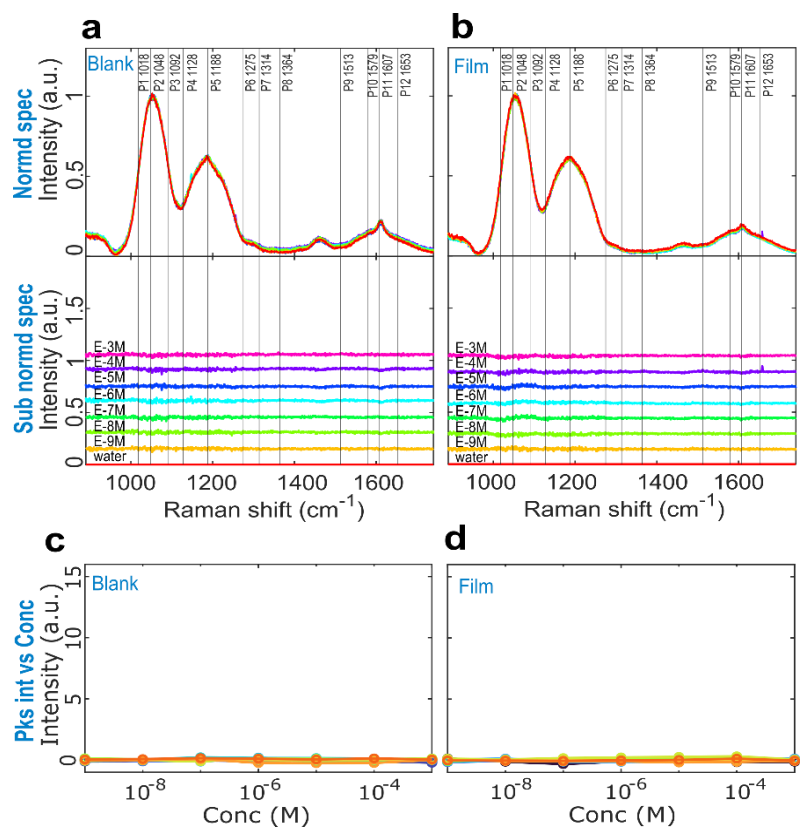

*Figure S6* The R6G LOD SERS spectra of blank and gold film-covered fibers. (a and b) The original spectra (up) and subtracted spectra (bottom) at different concentrations; the spectra were normalized to silica peak at 1055 $cm^{-1}$, and subtracted silica background (the silica background was taken as the normalized spectra measured in water). The spectra sets have been vertically offset for clarity. The vertical lines mark the 12 R6G peak positions. (c to d) The peak intensities (integrate peak areas) against the concentrations.

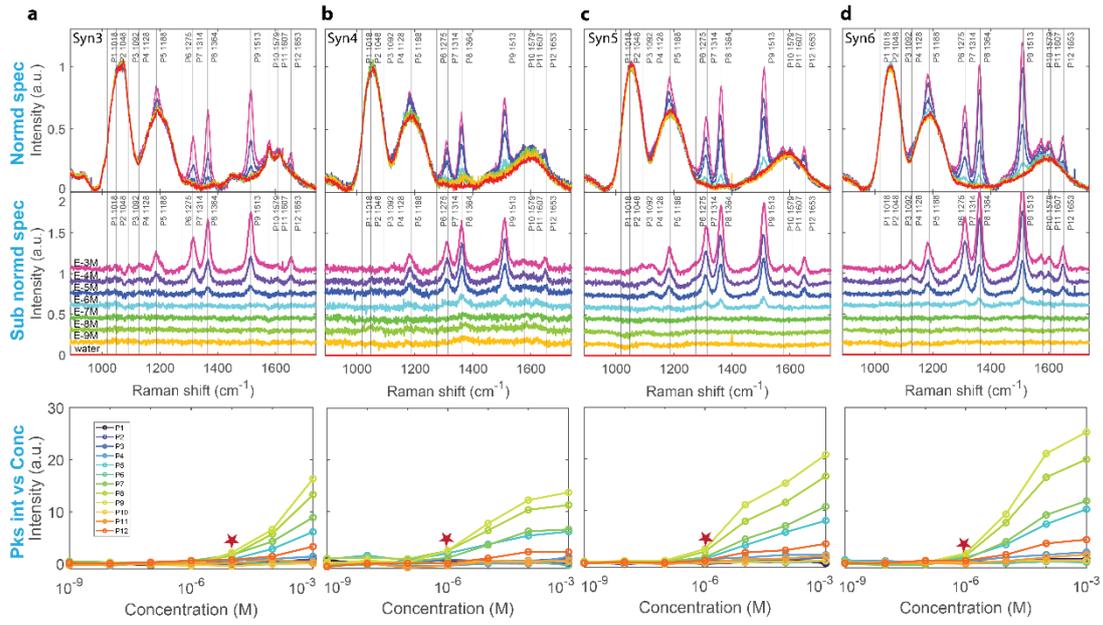

***Figure S7*** *(a-d) Representative SERS spectra of R6G LOD for Syn3-, Syn4-, Syn5-, and Syn6-HNI fibers. The upper panel displays the normalized original spectra and the background-subtracted spectra, while the bottom panel shows the peak intensities (integrated peak areas) as a function of concentration.*